\documentclass[a4paper,11pt]{article}
\usepackage{pos}
\usepackage{graphics}
\usepackage{subcaption}
\usepackage{pdfpages}

\usepackage{xspace}

\usepackage{lineno}

\newcommand{\epem}{e$^{+}$e$^{-}$\xspace}

\title{The CLIC potential for new physics}
\ShortTitle{The CLIC potential for new physics}

\manuallySeparateAuthors

\author*[1]{Jan Klamka}
\author{ on behalf of the CLICdp Collaboration}

\affiliation{Faculty of Physics, University of Warsaw\\
Pasteura 5, 02-093 Warsaw, Poland}

\note{Supported by the National Science Centre, Poland, project nr 2017/25/B/ST2/00496.}

\emailAdd{jan.klamka@fuw.edu.pl}

\abstract{The Compact Linear Collider (CLIC) is a mature option for a future
electron-positron collider operating at centre-of-mass energies of up
to 3\,TeV. 
It incorporates a novel two-beam acceleration technique offering
accelerating gradient of up to 100\,MeV/m. 
CLIC would be built and operated in a staged approach with
three centre-of-mass energy stages currently assumed to be 380\,GeV,
1.5\,TeV, and 3\,TeV. 
The first CLIC stage will be focused on precision Higgs and top quark measurements.
The so called ``Higgs-strahlung`` process (\epem $\to$ ZH) is a key for a model
independent measurement of Higgs boson decays and extraction of its couplings.
Precision top quark measurements will include the pair-production threshold scan, 
which is assumed to be the most precise method for the top-quark mass determination. 
The two subsequent energy stages will allow for extended Standard Model studies,
including the direct measurement of the Higgs self-coupling and the top Yukawa
coupling, but their main goals will be to search for signatures of 
Beyond the Standard Model phenomena. 
Presented in this contribution is a selection of recent results showing 
sensitivity of CLIC experiment to diverse BSM
physics scenarios. Compared with
hadron colliders, the low background conditions at CLIC provide
extended discovery potential, in particular for the production through
electroweak and/or Higgs boson interactions. This includes scenarios
with extended scalar sectors, also motivated by dark matter, which can
be searched for using associated production processes or cascade
decays involving electroweak gauge bosons. In a wide range of models, new particles
can be discovered almost up to the kinematic limit while the indirect search 
sensitivity extends up to ${\cal{O}}(100)$\,TeV scales.}

\FullConference{%
  *** The European Physical Society Conference on High Energy Physics (EPS-HEP2021), ***\\
  *** 26-30 July 2021 ***\\
  *** Online conference, jointly organized by Universität Hamburg and the research center DESY ***
}

\begin{document}

\maketitle

\section{Introduction}

Sensitivity to the new physics phenomena at the TeV scales is one of the main 
arguments for considering CLIC as the next energy-frontier machine.
While precision measurements will provide many indirect constraints on the Beyond Standard Model (BSM) physics already at the first CLIC stage, the main goals of CLIC will be to search for new physics directly, taking advantage of the lepton collider clear environment and availability of the simple signatures. This talk presents the latest studies of the CLIC sensitivity to the BSM physics.

\section{Dark matter searches in the mono-photon channel}

Mono-photon signature is the most general approach to the dark matter (DM) searches
at the \epem colliders. In this channel, pair-produced DM particles leave detector unobserved and the event can be tagged by an initial-state photon emission, which is precisely described by the Standard Model and only indirectly depends on the DM production.

\begin{figure}[bt]
	    \centering
	 	 \begin{subfigure}{0.56\textwidth}
	 	 	\centering
	 	 	\includegraphics[width=\linewidth]{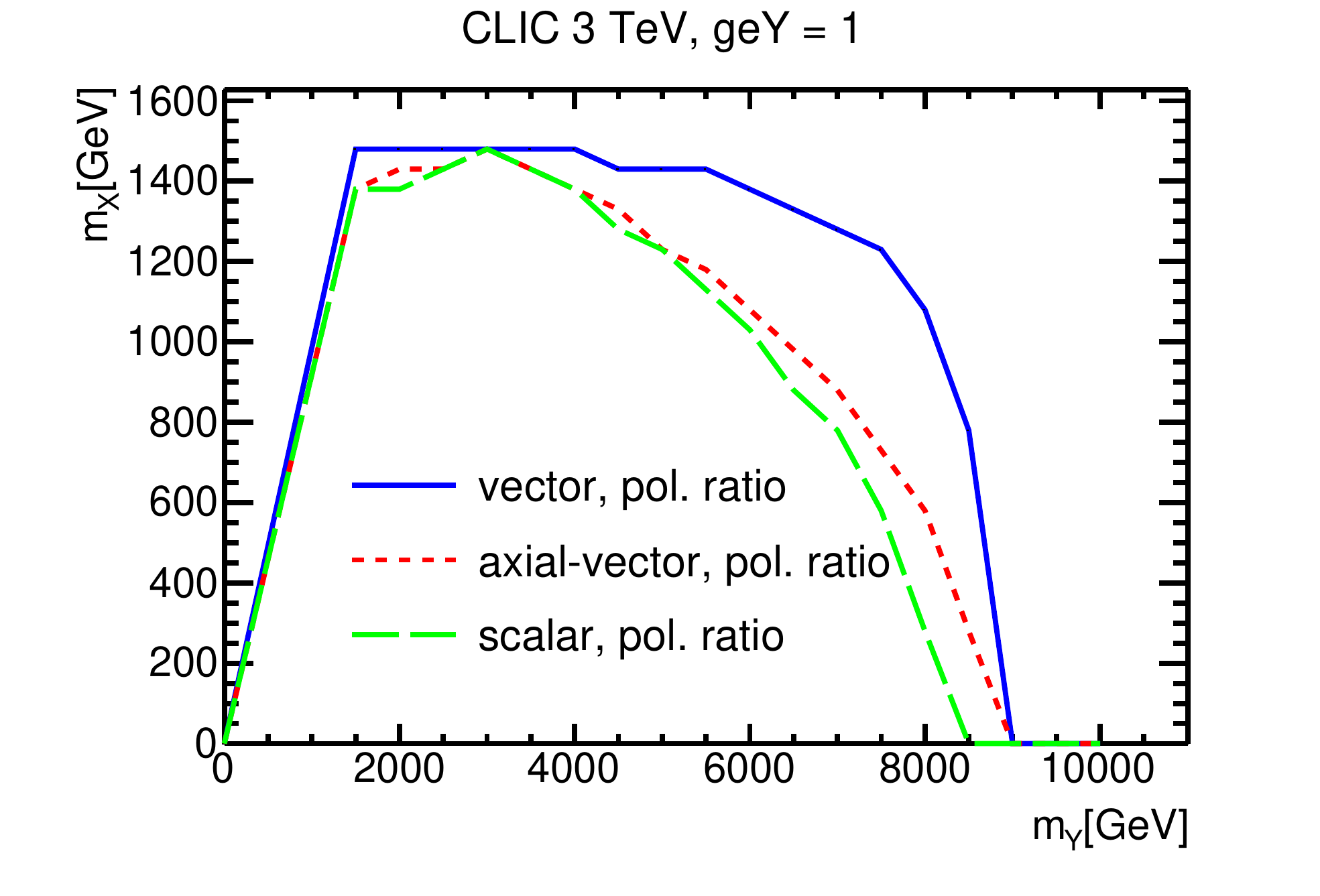}
	 	 \end{subfigure}%
	 	 \begin{subfigure}{0.44\textwidth}
	 	 	\centering
	 	 	\includegraphics[width=\linewidth]{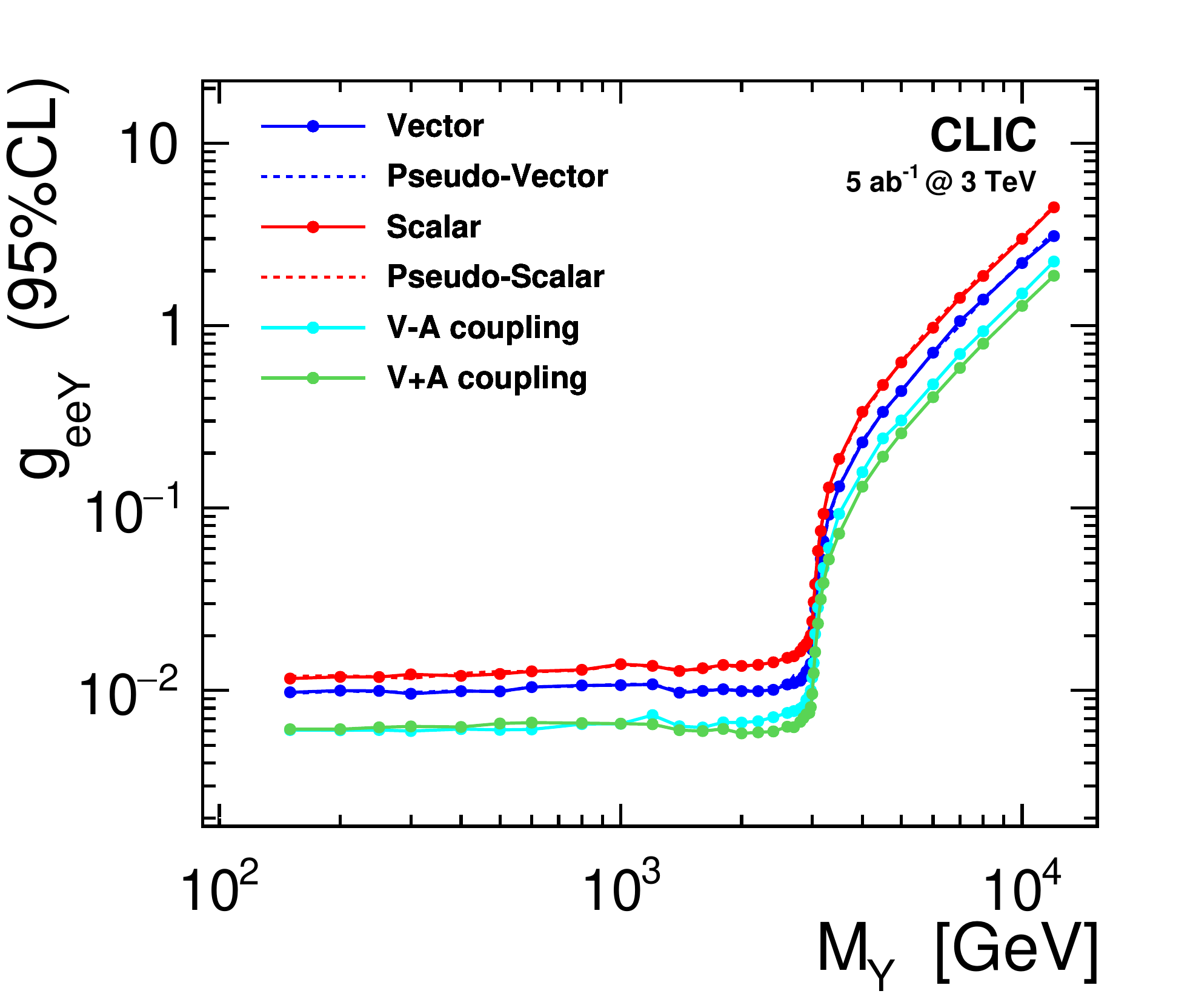}
	 	 \end{subfigure}
	 	 \caption{DM searches in the mono-photon channel at 3\,TeV CLIC. Left: 95\% C.L. exclusion limits on the mediator masses $m_Y$ and DM masses $m_X$, for $g_{eY}=1$ and three mediator types~\cite{Blaising:2021vhh}. Right: 95\% C.L. limits on the mediator coupling to electrons (denoted in \cite{Kalinowski:2021tyr} as $g_{eeY}$) as a function of the mediator mass (denoted in \cite{Kalinowski:2021tyr} as $M_Y$), for different mediator coupling structures, the relative mediator width $\Gamma/M=0.03$ and fermionic DM with mass of 50\,GeV~\cite{Kalinowski:2021tyr}. }
	 	 \label{fig:monophoton}
	 \end{figure}

The CLIC potential for DM detection has been studied for the 3\,TeV running scenario, considering large WIMP and mediator masses~\cite{Blaising:2021vhh}. 
CLIC sensitivity to the DM production is enhanced by the use 
of $\pm80$\,\% electron beam polarisation.
For the vector, axial-vector and scalar mediators considered in the Simplified Dark Matter Models framework, the best limits were obtained using the cross section ratio for the left-handed and the right-handed electron beam polarisation. 
The 95\% C.L. exclusion limits in the plane of mediator and DM particle masses, 
($m_Y$, $m_\chi$), are shown in figure~\ref{fig:monophoton} (left) for three types of the mediator coupling and the coupling value $g_{eY}=1$. 
The shape of the measured cross section ratio allows for the discrimination between vector and axial-vector model hypotheses and the WIMP mass can be determined with an accuracy of up to 1\%.

Scenarios with low mediator masses are not excluded, if the mediator coupling to SM particles is small. Sensitivity to different low mass scenarios with mono-photon signature at 3\,TeV CLIC was considered in a recent study~\cite{Kalinowski:2020lhp,Kalinowski:2021tyr}.
A novel ``experimental-like`` approach was proposed, where the expected exclusion limits are defined in terms of the DM pair-production cross section as a function of the mediator mass and width. In this approach, expected limits hardly depend on the assumed DM type and coupling structure. Expected 95\% C.L. limits on the mediator coupling to electrons are shown in figure~\ref{fig:monophoton} (right) as a function of the mediator mass and different mediator coupling scenarios. Results from combined analysis of data taken with different electron beam polarisations take into account systematic uncertainties from luminosity measurement, background normalisation, beam polarisation and beam energy spectra shape. For high mediator masses, limits on the mediator couplings were used to estimate the corresponding limits on the mass scale of new interactions in the EFT approach.  Limits range from 6.1\,TeV for scalar and 6.6\,TeV for vector mediator scenarios, to 10.1\,TeV for mediator with V+A coupling structure.

\section{Extended scalar sector}

Within the Standard Model, only about 0.1\% of the Higgs boson decays is ``invisible``, resulting from the ZZ$^*$ decay channel, where both Z boson decay into pair of neutrinos. However, increased branching fraction for the invisible channel is expected in many extensions of the SM, when Higgs decays into new,  weakly-interacting  particles that escape from the detector unobserved. Prospects for direct detection of invisible Higgs boson decays at CLIC have been studied for CLIC running at 380\,GeV~\cite{Mekala:2020zys}. 
The analysis was based on Whizard 2.7.0~\cite{Kilian:2007gr}  simulation of Higgs boson production and background processes and \textsc{Delphes} framework~\cite{deFavereau:2013fsa} for fast simulation of detector response. Expected CLIC beam spectra as well as  backgrounds resulting from $\gamma\gamma$ and $\gamma$e$^\pm$ hard interactions were taken into account.
Assuming that the measured event distributions are consistent with the predictions of the Standard Model, the 95\% C.L. limit on the invisible Higgs branching ratio expected at 380\,GeV CLIC is 1\% (0.5\%) for integrated luminosity of 1\,ab$^{-1}$ (4\,ab$^{-1}$).

\begin{figure}[bt]
	    \centering
	 	 \begin{subfigure}{0.49\textwidth}
	 	 	\centering
	 	 	\includegraphics[width=\linewidth]{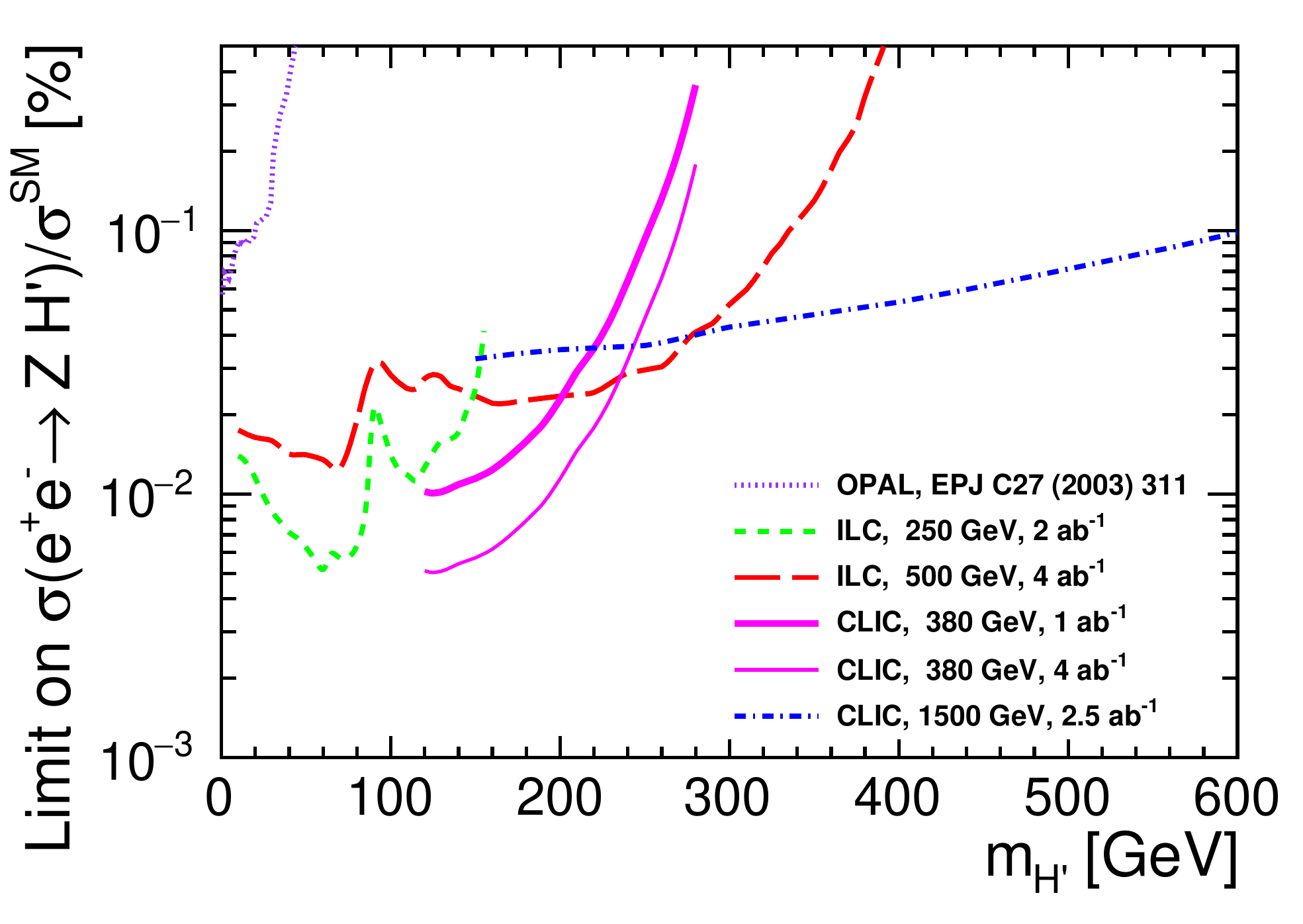}
	 	 \end{subfigure}%
	 	 \begin{subfigure}{0.51\textwidth}
	 	 	\centering
	 	 	\includegraphics[width=\linewidth]{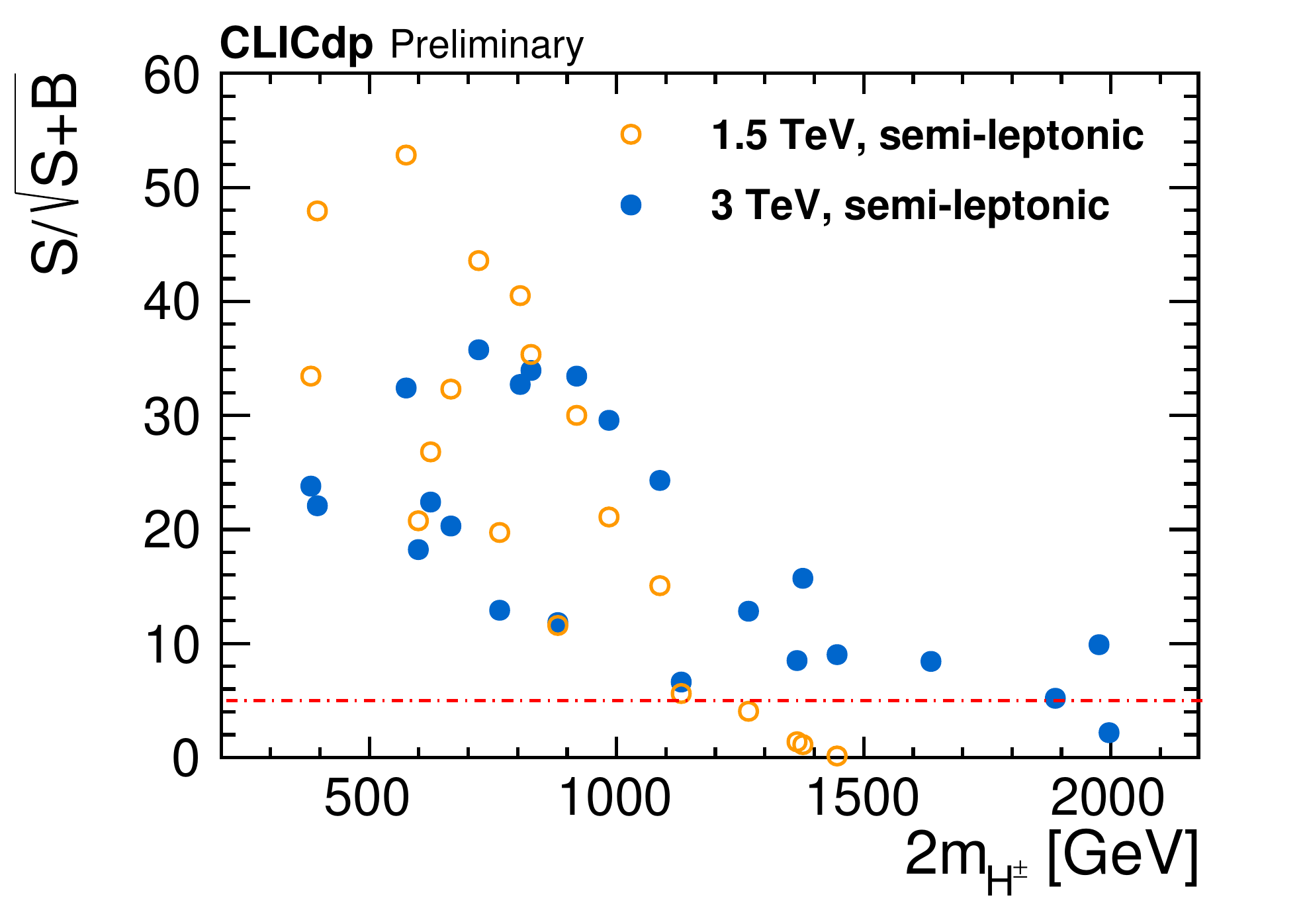}
	 	 \end{subfigure}
	 	 \caption{Discovery reach for the scalar sector extensions. Left: expected limits on the production of scalar H' with respect to the SM background, for 380\,GeV (1\,ab$^{-1}$ and 4\,ab$^{-1}$ option) and 1.5\,TeV CLIC running, compared to other \epem colliders. Right: statistical significance of deviations from the SM background, expected at 1.5\,TeV and 3\,TeV CLIC for the selected IDM benchmark scenarios. }
	 	 \label{fig:scalars}
	 \end{figure}

Limits on the invisible scalar decays can also be considered in the framework of the Higgs portal models, where another scalar, H', mixing with the SM Higgs boson, could be produced in \epem collisions in a process similar to the SM Higgs-strahlung channel: \epem$\to$ ZH'. Expected limits on the production cross section for this process, relative to the SM cross section for the given Higgs boson mass, are shown in figure~\ref{fig:scalars} (left) as a function of the new scalar mass, $m_{H'}$. The results are shown for CLIC running at 380\,GeV, with two luminosity scenarios, and for the 1.5\,TeV CLIC stage, compared with other \epem colliders. These limits, when interpreted in terms of the Vector Fermion Dark Matter Model, correspond to limits on the scalar mixing angle much stronger than the current constraints from ATLAS, CMS and indirect measurements.

Production of DM particles at CLIC was also studied in the framework of the Inert Doublet Model (IDM), one of the simplest extensions of the SM, where just one additional doublet is introduced. It contains four new scalar fields, H$^\pm$, A and H, from which the latter is stable, what makes it a natural DM candidate. The CLIC potential for observing IDM scalar production in the leptonic channels is limited by small branching ratios~\cite{Kalinowski:2018kdn}.
Much higher sensitivity is expected using the semi-leptonic final state~\cite{Klamka:2728552}, which offers higher statistics. A set of 23 benchmark points \cite{Kalinowski:2018ylg} in the model five-dimensional parameter space was considered. For the five selected scenarios, full simulation of detector response was used and, to extend the scope of the study, \textsc{Delphes} toolkit provided fast simulation for all of the considered benchmarks.

For the IDM scenarios with small scalar mass splitting, reconstruction of the low-energetic final state is highly influenced by the background coming from beam-induced $\gamma\gamma$ interactions producing soft hadrons, the so called \textit{overlay events}.  The standard mitigation of this background at CLIC is achieved by applying timing cuts on the reconstructed particles. As these cannot be implemented in the CLICdet model for \textsc{Delphes}, the overlay events were included in fast simulation using approximate cuts applied on the generator level to the $\gamma\gamma\to$ hadrons samples. With this procedure,  good agreement between the fast and full simulation results is obtained~\cite{Klamka:2021vqp}. In figure~\ref{fig:scalars} (right) the expected statistical significance of IDM scalar production observation is presented as a function of the charged inert scalar mass, $m_{H^\pm}$. Shown are results obtained using \textsc{Delphes} simulation for all benchmark scenarios considered in the study, both at 1.5\,TeV and 3\,TeV CLIC running stages. CLIC sensitivity to charged IDM scalar production is significantly increased with respect to the previous study \cite{Kalinowski:2018kdn} and indicate that the inert scalars could be discovered at CLIC in almost all of the considered model scenarios, with significance reaching up to~50\,$\sigma$.

\section{Heavy neutrinos}

Many problems of the SM, including the nature of dark matter or CP violation, can be related to the neutrino sector, which is not well constrained. We do not understand the very nature of neutrinos or the origin of their masses. All these problems can be solved by introducing the additional neutrinos (or ``neutral leptons``), that in principle could be heavy and might be produced in particle colliders. 

\begin{figure}[bt]
	    \centering
	 	 	\includegraphics[width=0.6\textwidth]{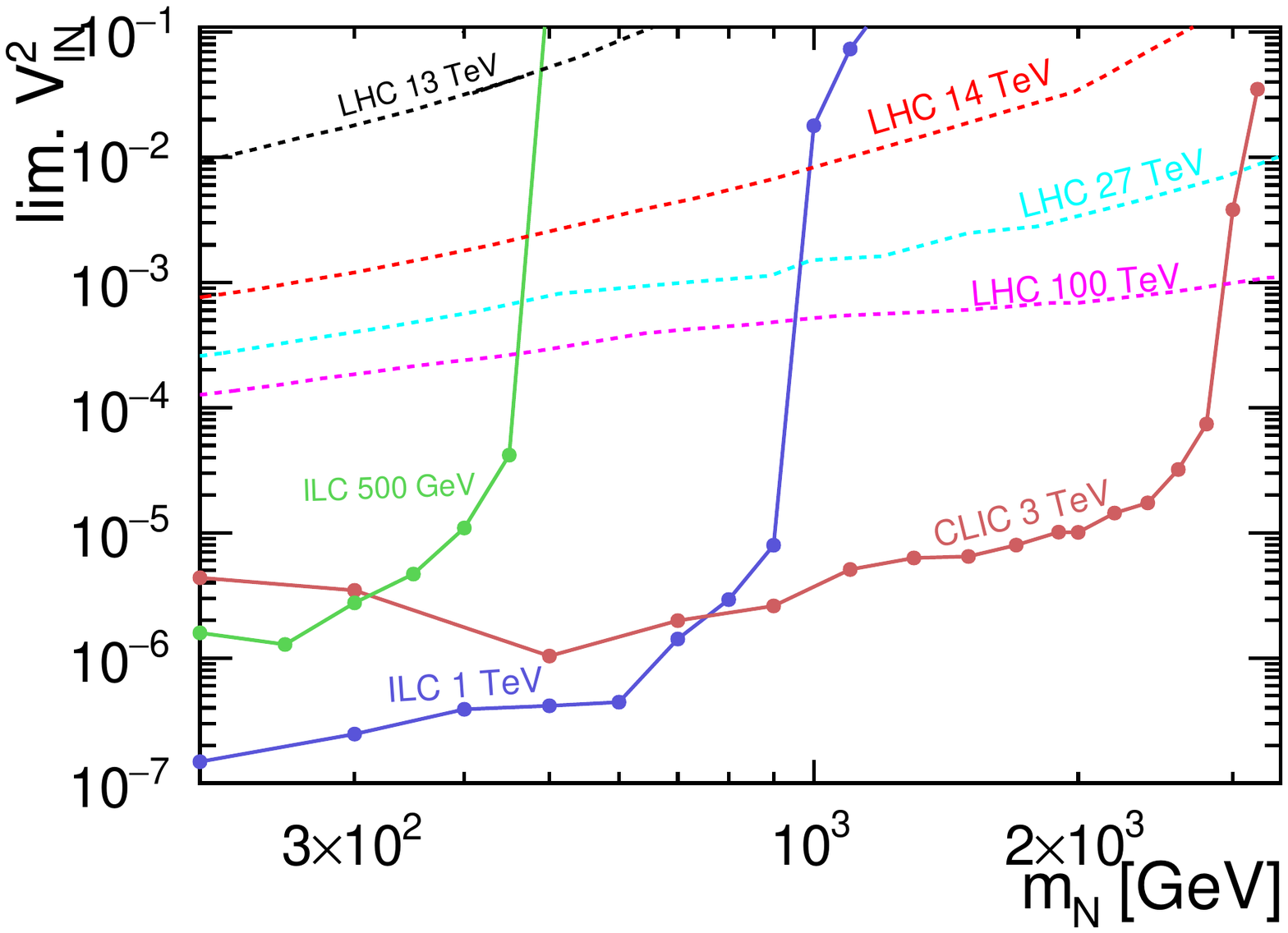}
	 	 \caption{Expected 3\,TeV CLIC sensitivity to coupling $V^{2}_{\mathrm{lN}}$ of heavy neutrino to SM leptons, compared with the expected limits from ILC and hadron colliders. }
	 	 \label{fig:neutrinos}
	 \end{figure}

A possibility of detecting heavy neutrino production at CLIC was studied for a model with additional right-handed massive neutrinos~\cite{Pascoli:2018heg}. Considered was the process of light-heavy neutrino pair production with the semi-leptonic final state allowing for full reconstruction of the new particle. Both Dirac and Majorana scenarios were studied, with masses in range 200-3200\,GeV~\cite{Mekala:PoS}. Signal events, as well as backgrounds from \epem, $\gamma\gamma$ and $\gamma$e$^\pm$ hard interactions (both from the beamstrahlung and the Equivalent Photon Approximation), were simulated in Whizard 2.8.5. The analysis was based on fast detector response simulation with \textsc{Delphes} and the use of Boosted Decision Trees for event classification.  Expected limits on the heavy neutrino production cross section were evaluated using CLs method and translated into constraints on the coupling parameter $V_{lN}^2$ describing heavy neutrino coupling to SM leptons (assumed to be independent of the lepton flavour). Results are presented in figure~\ref{fig:neutrinos} as a function of the neutrino mass, $m_N$, and compared with the corresponding predictions for the ILC and different hadron collider options (all concern Dirac neutrinos, except for current limits from the LHC). Presented results confirm that the sensitivity of CLIC to heavy neutrino production extends up to the kinematic limit, is comparable to the ILC (with higher mass reach) and much better than expected in the current or future hadron colliders.

\section{Conclusions}

\begin{figure}[bt]
	    \centering
	 	\includegraphics[width=0.8\textwidth]{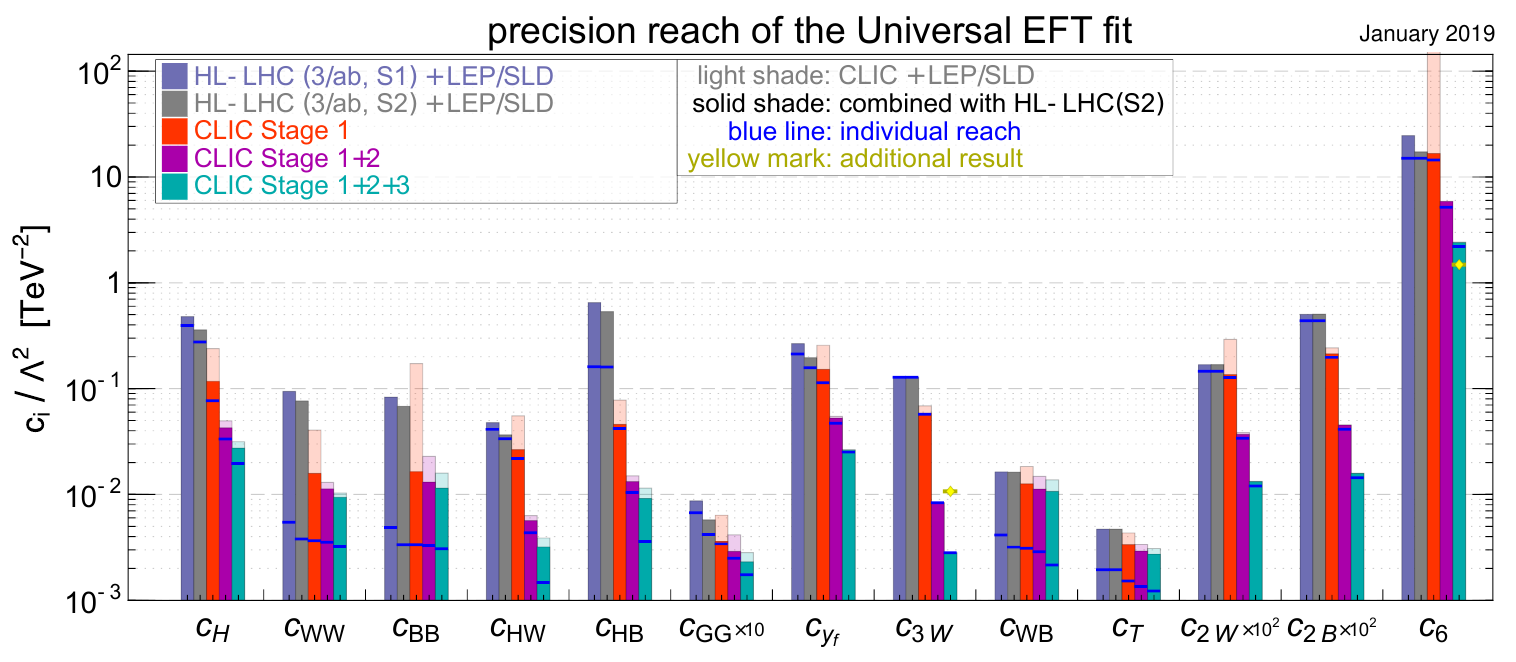}
	  \caption{Results of a global fit of universal EFT operators to precision
	    Higgs, top and EW observables at CLIC \cite{deBlas:2018mhx}. }
	 	 \label{fig:eft}
	 \end{figure}

With the clean environment of \epem collisions and its high energy reach, CLIC offers a great possibility of direct searches for BSM physics. Presented results indicate that in many cases it surpasses not only the expected capabilities of the HL-LHC, but can also be competitive to future hadron colliders such as FCC-hh or to the ILC project. Apart from the presented direct searches, precision measurements at CLIC probe new physics scenarios extending up to ${\cal{O}}(100)$\,TeV scales. Example limits on the new physics effects extracted in the EFT framework are presented in figure~\ref{fig:eft} \cite{deBlas:2018mhx}. All direct and indirect search opportunities make CLIC a fantastic future collider option.

%

\bibliographystyle{JHEP}
\bibliography{eps-hep_clic.bib}

\end{document}